\documentclass[12pt]{iopart}

\usepackage{graphicx}
\usepackage{color}
\usepackage{cite}
\usepackage{bm}
\usepackage{amssymb}
\usepackage{amsfonts}
\renewcommand{\d}{{\rm d}}
\renewcommand{\i}{{\rm i}}
\newcommand {\ee}{{\rm e}}

\newcommand{\lamu}{\lambda_{\rm u}}
\newcommand {\E}  {\varepsilon}
\newcommand {\om} {\omega}
\newcommand {\Om} {\Omega}

\newcommand {\bfn} {{\bf n}}
\newcommand {\bfp} {{\bf p}}

\newcommand {\bfr} {{\bf r}}
\newcommand {\bfv} {{\bf v}}

\newcommand {\bfE} {{\bf E}}

\newcommand {\bfR} {{\bf R}}

\newcommand {\bfrho} {\boldsymbol{ \rho}}

\begin{document}

\title[Channeling in a diamond crystal]
{Channeling of ultrarelativistic particles in a diamond crystal}

\author{K.~B. Agapev$^{1}$, V.~K. Ivanov$^1$,  A.~V. Korol$^{2}$, 
A.~V. Solov'yov$^2$\footnote{On leave 
from A.F. Ioffe Physical-Technical Institute, St. Petersburg, Russia}}

\address{$^1$ Peter the Great St. Petersburg Polytechnic University, 
Polytechnicheskaya 29, 195251 St. Petersburg, Russia}
\address{$^2$ MBN Research Center, Altenh\"{o}ferallee 3, 
60438 Frankfurt am Main, Germany}

\ead{korol@mbnexplorer.com}

\begin{abstract}
The results of numerical simulation of the channeling of ultrarelativistic 270 MeV electrons
and positrons in a diamond crystal are presented.
Using the MBN Explorer package \cite{MBN_Explorer_Paper,MBN_Explorer_Site}, the trajectories of the have been determined of the particles’ incident on
a 20 $\mu$m thick crystal along (110) crystallographic plane. 
The channeling parameters and radiation spectra of electrons and positrons have been computed for the cases of 
straight and periodically bent diamond crystals.
\end{abstract}



\section{Introduction \label{Introduction}}

The processes of interaction of charged particles with matter, in particular, crystalline
solids, have been long studied both experimentally and theoretically. 
The goal of these studies is to determine such characteristics of the interaction as the mean free path traveled
by particles in the material, their energy losses, emission spectra, and others.

Channeling in crystals, when charged particles falling into a potential channel shaped by electrostatic forces propagate along
crystallographic planes or axes, has become the focus of much attention in recent years.
The particles trapped in a channel of a straight crystal can travel long distances exceeding the mean free path in an 
amorphous target, since such particles lose considerably less energy along their path \cite{Lindhard_KDan_v34_p1_1965}. 
For electrons, the channel lies along atomic rows or ion chains of the crystal, while for positrons it lies in the space between
atomic rows. 
The stability of particle motion along the channels depends on the energy of the transverse motion that is low compared
with the height of the potential barrier.

A particle trapped in the channel experiences oscillations in a plane transverse
to the direction of the particle’s propagation, inducing radiation during its channeling \cite{ChRad:Kumakhov1976}.
This radiation is determined by the transverse energy of the channeled particle, and its intensity
varies depending on the type of crystal and its orientation. 
Oscillatory radiation is incoherent and has a broad energy spectrum
\cite{ChRad:AndersenEtAl1983,BakEtAl_NPB_v254_p491_1985,
BakEtAl_NPB_v525_p302_1988,BazylevZhevago:Uspekhi-v28-p565-1982,KumakhovKomarov-AIP}.

Channeling can also occur in bent crystals, which are often used to bend charged particle
beams accelerated to relativistic energies \cite{Tsyganov_TM-682_1976}. 
The motion of a particle consists of two components: its oscillatory motion in the
channel and its propagation along the centerline of the bent channel. 
The stability of the second component of motion in such a bent channel is provided by an additional condition, namely,
that the bending radius $R$ should significantly exceed the critical value $R_c$ determined by the
energy of the particle  \cite{Tsyganov_TM-682_1976}. 
This motion of a relativistic particle trapped in a bent channel induces additional synchrotron radiation. 
The intensity and frequency of synchrotron radiation depend on the type and energy of the channeled particles, as well as on
the characteristics of the crystal \cite{KaplinVorobev1978,Bashmakov1981,TaratinVorobiev1989,ArutyunovEtAl_NP_1991,
Taratin_PhysPartNucl_v29_p1063_1998-English,KSG1998,KSG_review_1999,ChannelingBook2014}.

Undulator radiation is certainly an interesting subject to explore in connection with the concept of the crystal undulator (see,
for example, Ref. \cite{ChannelingBook2014} and references therein). 
Channeling of charged relativistic particles in a periodically bent crystal (a crystal undulator)
can produce a new source of monochromatic radiation with energies ranging from hundreds
of keV to several MeV.

There has been a number of experiments in the recent years with a view to create the crystal
undulator, measuring the channeling parameters and the characteristics of the emission spectra
of ultrarelativistic positrons \cite{BaranovEtAl_CU_2006,Backe_EtAl_NuovoCimC_v34_p175_2011,Backe_EtAl_2008}
and electrons \cite{Backe_EtAl_2011,Backe_EtAl_2013} 
in straight and bent crystals of silicon and diamond. 
Theoretical studies on channeling in these crystals are carried out using the newly developed MBN Explorer
package \cite{MBN_Explorer_Paper,MBN_Explorer_Site}. 
Simulations for amorphous and crystalline silicon have verified that this package is applicable for describing 
the channeling of electrons and positrons 
\cite{MBN_ChannelingPaper_2013,Sub_GeV_2013,PolozkovEtAl:NTV_v1_p212_2015,Korol_EtAl_NIMB_v387_p41_2016}. 

Since experiments are currently being carried out to measure the emission spectra of electrons in a periodically bent 
diamond crystal \cite{BadEms_p58}, theoretical interpretation of the experimental results is clearly an interesting problem.

In view of the above, the goal of this study is theoretical analysis of channeling of ultrarelativistic 
electrons and positrons with an energy of 270 MeV both in a straight diamond crystal oriented along the (110) 
crystallographic plane and in a periodically bent diamond crystal.

We have performed simulations of electron and positron channeling in straight, bent and
periodically bent channels using the versatile MBN Explorer software package.

\section{Simulation procedure with the MBN Explorer package \label{Procedure}}

Three-dimensional simulation of ultrarelativistic particles passing through a
crystalline medium is carried out using a molecular dynamics algorithm implemented in
the MBN Explorer software package \cite{MBN_ChannelingPaper_2013}. 
The characteristics of the motion of high-energy particles inside the crystal were obtained by
integrating the relativistic equations of motion.
Step-by-step dynamic simulation of the crystalline medium was performed to construct
the particle trajectory.

A quasi-classical approximation is applicable to describing the motion of ultrarelativistic
particles, and, since the quantum corrections are small, it is sufficient to use the equations of
classical relativistic mechanics:
\begin{eqnarray}
\dot{\bfp} = q \bfE(\bfr)
\label{eq.01}
\end{eqnarray}
Here $\bfE(\bfr)$ is the external electrostatic field, 
$q$ is the particle charge, and $\bfp$ is its relativistic
momentum $\bfp = m\gamma \bfv$,
where $m$ and $v$ are the mass and velocity of the particle, respectively,
$\gamma = \left(1 -v^2/c^2\right)^{-1/2} \gg1 $  is
the relativistic factor  ($c$ is the speed of light).

Initial values of the coordinates and velocity of the particle
are used to integrate Eq. (\ref{eq.01}).

In the MBN Explorer channeling module, the force $q \bfE(\bfr)$ is calculated as the gradient of 
the electrostatic potential $U(\bfr)$ equal to the sum of atomic potentials $U_{\rm at}$:
\begin{eqnarray}
U(\bfr) = \sum_{j} U_{\rm at}(\bfrho)
\label{eq.02}
\end{eqnarray}
where $\bfrho_j = \bfr - \bfR_j$ with $\bfR_j$ standing for the position vector of a $j$th atom.

Formally, the sum in (\ref{eq.02}) accounts for all crystal atoms. 
However, given a rapid decrease of $U_{\rm at}(\bfrho)$  with distance, one can introduce the maximum 
distance $\rho_{\max}$, beyond which the contribution of the atomic potential is negligible. 
Therefore, for a given observation point $\bfr$, the sum can be limited to the atoms located inside a 
sphere with the radius $\rho_{\max}$. 
The linked cell algorithm implemented in the MBN Explorer is used to search for such atoms. 
This algorithm involves dividing the crystal into cells and considering only
the atoms closest to the particle. 
The described scheme is used to calculate the force $q \bfE(\bfr)$ acting on the projectile 
at each step of integration.

The motion of particles along a crystallographic plane with the Miller indices
$(k l m)$ is simulated by the following procedure \cite{ChannelingBook2014,MBN_ChannelingPaper_2013}. 
A simulation box with the dimensions $L_x\times L_y \times L_z$ 
is introduced, containing a crystal lattice. 
The $z$ axis is oriented along the incident beam and is parallel to the
$(k l m)$ plane, the $y$ axis is perpendicular to this plane. 
The position vectors of the lattice sites are generated in accordance with the type of the
Bravais cell of the crystal, using predefined values of ​​ the translation vectors.

Once the nodes inside the simulation box are determined, the position vectors of the
atomic nuclei are generated taking into account the thermal vibrations of these nuclei
resulting in a random displacement from the nodal positions; these displacements are determined
by the normal distribution with respect to the root-mean-square amplitude of thermal vibrations \cite{Gemmel}.

Integration of the equations of motion begins at instant $t = 0$, when the particle enters the crystal 
at $z = 0$.
A random number generator is used to choose 
The initial transverse coordinates $x_0$ and $y_0$ are generated randomly.
For a beam with zero emittance, the initial velocity $\bfv_0$ is oriented along the $z$ axis.
Particle propagation through a crystal with a finite thickness $L$ is simulated in MBN Explorer using 
the so-called dynamic simulation box \cite{ChannelingBook2014,MBN_ChannelingPaper_2013} as a new type of boundary conditions. 
A particle moving inside the box interacts with atoms lying inside the cutoff sphere.
To optimize the numerical procedure, the dimensions of the box are chosen to be
3 to 5 times larger than $\rho_{\max}$.
At the instant when the distance $l$ from the particle to the nearest face of the box becomes close to $\rho_{\max}$, 
a new simulation box of the same size is generated, with its geometric center approximately coinciding with the position of
the particle. 
The atoms located at the intersection of the old and the new simulation boxes are left intact.
The positions of the atoms in the rest of the new box are generated anew.
Simulation is interrupted when the $z$ coordinate of the particle
becomes equal to the crystal thickness $L$.

\section{Simulation of electron and positron trajectories \label{Trajectories}}

The MBN Explorer package was used to simulate the trajectories of 270 MeV electrons
and positrons incident on diamond crystals along the (110) crystallographic planes. 
The calculations were performed for a straight crystal and for a crystal with periodical cosine-like bending. 
In both cases the crystal length was set to $L=20$ $\mu$m. 
The periodical bending was considered with the amplitude $a=2.5$ \AA{} and 
period $\lamu=5$ $\mu$m. 
Each set of calculations included simulation of  6000 trajectories of projectiles 
which were analyzed further to calculate the channeling parameters and radiation emission.

An ordinary diamond crystal has straight channels due to the periodic arrangement of its atoms. 
The width of the channel is determined by the interatomic distance which is $d = 1.26$\AA{}. 
Particles trapped in straight channels with a low transverse energy leave such
channels less often. 
Since the crystal is short enough, positrons most often move through the entire
straight crystal while staying in the channel, and electrons are more likely to collide with
lattice atoms and leave the channel. 
This is because positrons move between the crystal atoms, where they are confined by repulsive
interaction with the lattice ions. 
On the other hand, electrons move along helical trajectories in the immediate vicinity of the nuclei, so they
are much more likely to collide with them and escape the channel.

\begin{figure} [h]
\centering
\includegraphics[width=7.7cm,clip]{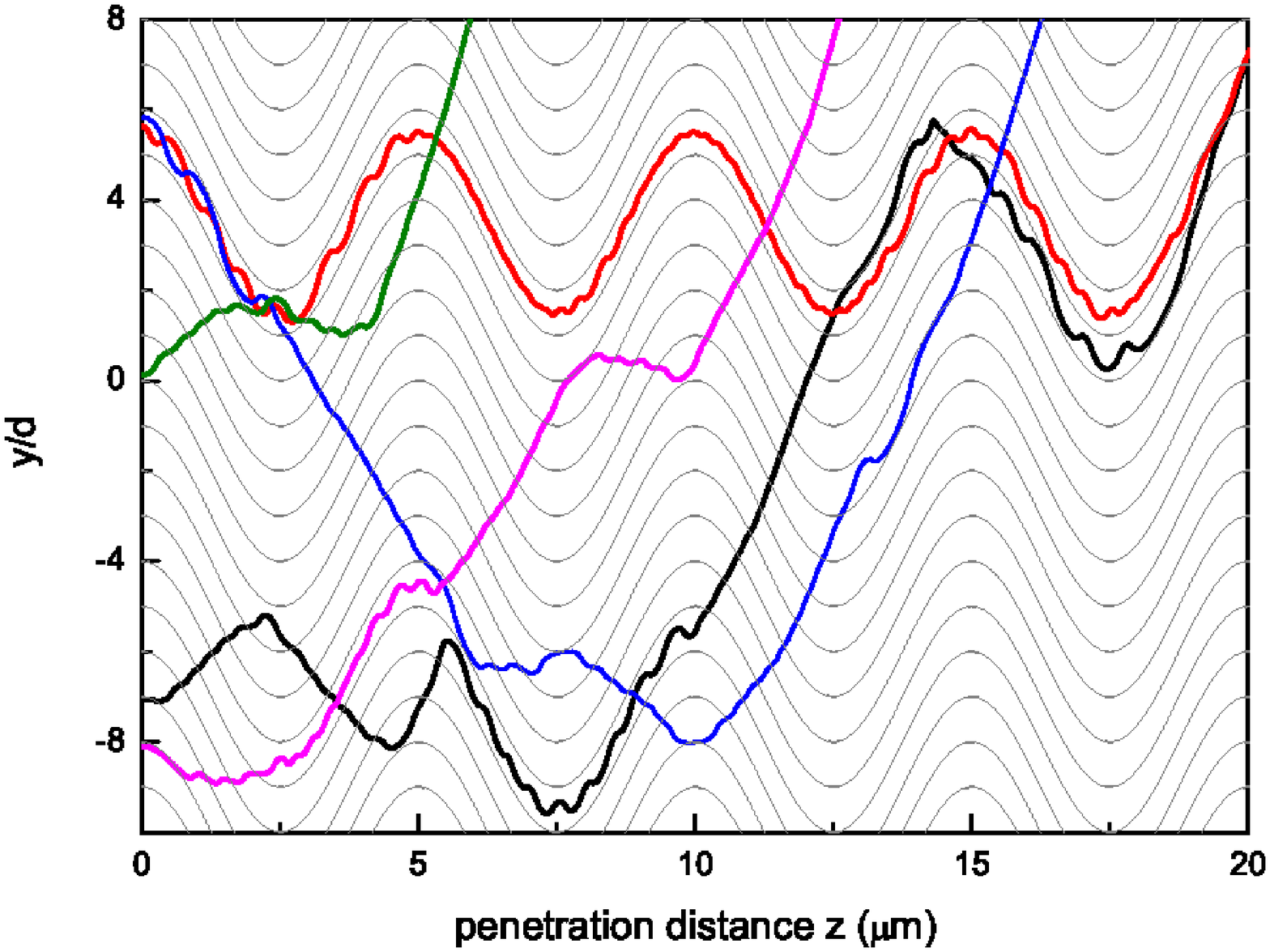}
\includegraphics[width=7.7cm,clip]{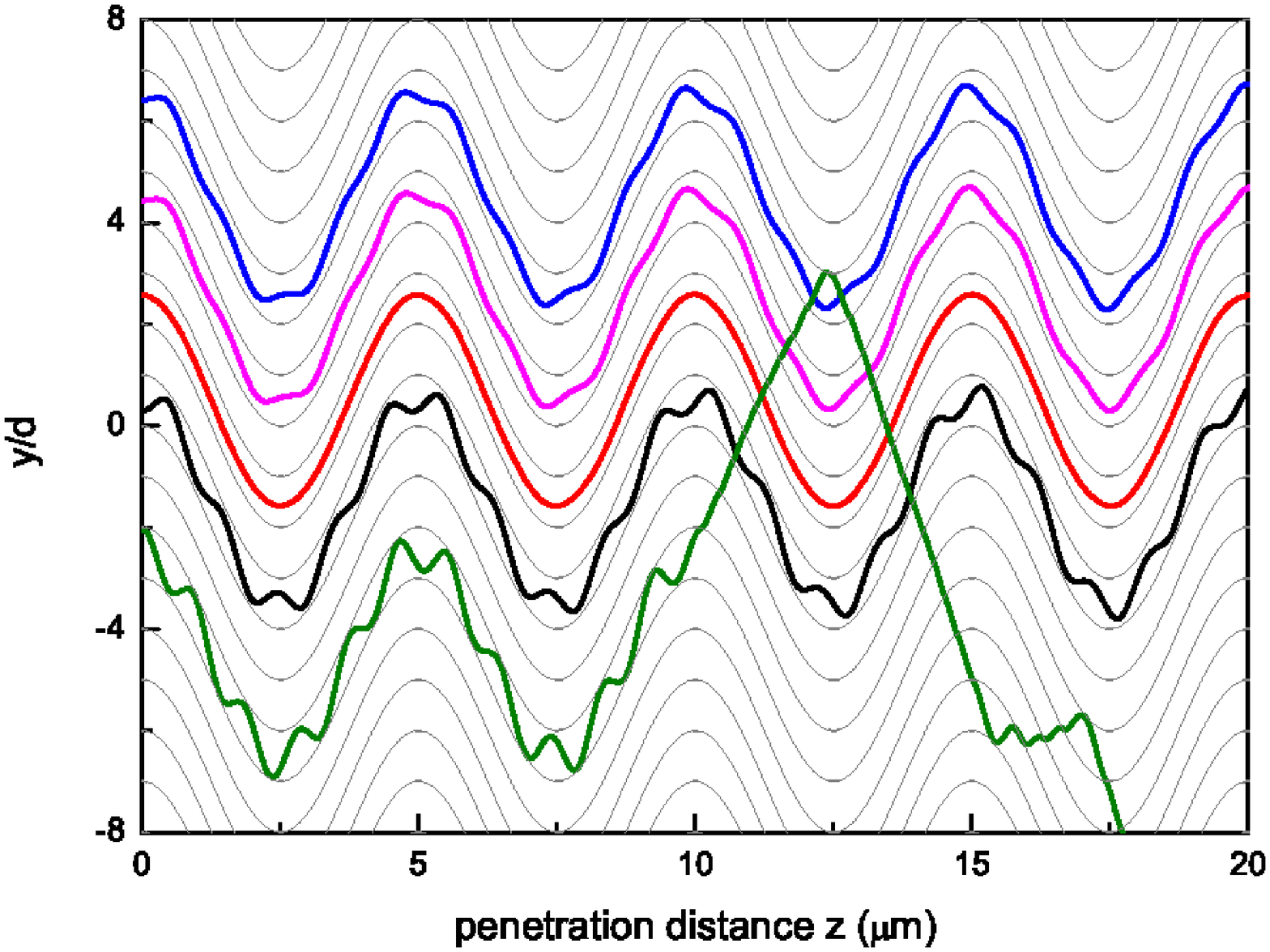}
\caption{
Representative trajectories of electrons (left) and positrons (right) with energies of 270 MeV
in a periodically bent 20 $\mu$m thick oriented diamond(110) crystal. 
Channeling (curves 1), dechanneling (2) and rechanneling (3) modes are indicated.
}
\label{Figure1.fig}
\end{figure}

The trajectories of charged particles channeled in bent crystals become more complex and diverse. 
As an example, Fig. \ref{Figure1.fig} shows several typical trajectories of electrons (left panel) 
and  positrons (right panel) in periodically bent diamond. 
Thin solid lines in the figure indicate the boundaries of the channels; the distance
$y$ is plotted along the vertical axis in a plane perpendicular to the direction of motion
(the distance is measured in units of the interatomic spacing $d$). 
The main features and characteristics of particle motion in a crystal, such as the channeling, 
dechanneling, and rechanneling modes, are shown in the figures. 
Rechanneling is a process when a particle moving outside a channel can experience a
collision and get trapped into some channel as a result.

Figure \ref{Figure1.fig} left presents the trajectory of the only electron that propagated through 
a crystal staying in the same channel. 
Statistically, such trajectories are an exception, as the rest of the trajectories
presented correspond to the more typical motion of electrons in dechanneling and
irregular rechanneling modes in short segments of different channels.
Comparison of the trajectories shown Fig. 1 left and right, indicates that positrons channel
much better than electrons, and this pattern is observed for both straight and bent crystals.
Only a small part of the positrons originally trapped in the channel escapes it, while most
of them move through the entire crystal while staying in one channel.
Therefore, the intensity of synchrotron radiation should be higher in periodically bent crystal.

Notably, positrons may have different oscillation amplitudes inside the channel,
but transverse oscillations are practically isochronous and their period remains almost
unchanged, which corresponds to harmonic oscillations. 
Consequently, all positrons emit energy at approximately the same wavelength,
and their channeling radiation peak is narrower and more intense, in contrast to the maximum
radiation intensity for electrons.

Statistical analysis of the calculated trajectories allowed to obtain the main
parameters characterizing the channeling of charged particles (given in the table).

The particle trapping coefficient $A$ (acceptance) is the ratio of the number $N_{\rm acc}$
of the particles trapped in the channel upon entering the crystal to the number $N_0$ of all
incident particles: $A=N_{\rm acc}/N_0$.

The values given in the table refer to the acceptance for the particles falling along the $z$
axis.
The remaining parameters are related to the mean distances or the times during which
the charged particles stay in one or several channels. 
The channeling length $L_{\rm ch}$ is defined as the mean total distance traveled by a particle in
the channeling mode throughout the crystal. 
The rechanneling length $L_{\rm rech}$ is the mean distance covered by a particle from the moment 
when it dechannels until the opposite event of rechanneling, i.e. capture into the channeling mode
as a result of collisions with the crystal atoms.
To more parameters are listed in the Table.
These are so-called penetration lengths \cite{ChannelingBook2014,MBN_ChannelingPaper_2013}. 
The first one, denoted as $L_{\rm p1}$, is the mean distance traveled by a particle, accepted into the 
channeling mode at the entrance, until it dechannels at some point in the bulk.
The penetration length $L_{\rm p2}$ is calculated as the arithmetic mean of all channeling segments 
(initial and secondary) with respect to the total number channeling segments in all simulated trajectories.
Thus, it characterizes the average distance traveled by a particle in the channeling mode.

\begin{table}
\caption{
Channeling parameters of $855$ MeV positrons ($e^+$) and electrons ($e^-$) in
straight and periodically bent (PB) $20$ $\mu$m thick oriented diamond(110) crystal:
acceptance $A$, 
channeling length $L_{\rm ch}$,
rechanneling length $L_{\rm rech}$,
penetration lengths $L_{\rm p1}$ and $L_{\rm p2}$
(all in $\mu$m). 
}
\footnotesize\rm
\begin{tabular}{@{}rrrrrrr}
\br
Parameter      & \multicolumn{2}{c}{straight crystal}& \ & \multicolumn{2}{c}{PB crystal}\\ 
               &     $e^-$     &    $e^+$            & \ &    $e^-$     &    $e^+$        \\
\br  
 $A$           & 0.70          &  0.96               & \ &    0.51      &   0.89          \\
$L_{\rm ch}$   & 9.04          &  18.7               & \ &    6.06      &  17.2           \\
$L_{\rm rech}$ & 4.18          &  6.08               & \ &    5.98      &  7.53           \\
$L_{\rm p1}$   & 5.43          & 19.1                & \ &    4.30      & 18.8            \\
$L_{\rm p2}$   & 4.55          & 18.0                & \ &    3.60      & 16.4            \\
\br  
\end{tabular}
\label{Table_ep-data.C}
\end{table}

Since the crystal is rather short (20 $\mu$m), the positrons accepted in the channeling mode travel through
almost the entire crystal staying in the same channel, and, thus, they have greater penetration,
channeling and rechanneling lengths.
Electrons experience collisions with lattice ions at a higher rate, since their trajectories
pass in the immediate vicinity of the ions, and thus the dechanneling events are more frequent.

\section{Emission spectra of electrons and positrons}

For each projectile, the simulated dependences $\bfr = \bfr(t)$ and $\bfv = \bfv(t)$ 
allow one to calculate the spectral characteristics of the radiation emitted by the particle.

The spectral angular distribution of the radiated energy $\d^3 E / (\d\hbar\om \d \Om)$ 
($\om$ and $\Om$ stand for the frequency of radiation and the emission solid angle, respectively) is
calculated following the general formula derived within the quasi-classical approximation \cite{Baier}:
\begin{eqnarray} 
\fl
{\d^3 E \over \hbar\d\om\, \d \Om}
=
\alpha \,
{ q^2\omega^2  \over 8\pi^2 }
\int\limits_{-\infty}^{\infty} \d t_1\!
\int\limits_{-\infty}^{\infty} \d t_2\,
\ee^{\i \,\omega^{\prime} \left(\psi(t_1) -\psi(t_2)\right)}
\left[
\left( 1+(1+u)^2 \right)
\left(
{\bfv_1\cdot\bfv_2 \over c^2}  -1
\right)
+{u^2 \over \gamma^2}
\right]\,.
\label{eq.03} 
\end{eqnarray}
Here $\alpha= e^2/ \hbar\, c$ is the fine structure constant,
$q$ is measured in units of the elementary charge, $\bfv_{1,2} =\bfv(t_{1,2})$, 
and
the $\psi(t) = t - \bfn\cdot\bfr(t)/ c$, with $\bfn$ being the unit vector in the 
direction of radiation emission.
Other quantities, which account for the radiative recoil, are as follows:
$\om^{\prime} = (1+u)\, \om$ and $u = \hbar \om/(\E - \hbar \om)$.

For each individual trajectory $j$, the spectral distribution is calculated by 
numerically integrating the values of $\d^3 E_j / (\d\hbar\om \d \Om)$ 
over the ranges $\phi=[0,2\pi]$ and $\theta=[0,\theta_0]$, where 
$\theta_0$ is related to the detector aperture.
The resulting distribution is calculated averaging $\d^3 E_j$  
over the ensemble of the trajectories.

The results presented below refer to the emission within the cone
$\theta_0 \leq 0.2$ mrad. 

Figure 2 left shows the emission spectra of electrons in the straight and periodically bent crystal. 
The broad peak (curve 1) at $\hbar \om \geq 0.4$ MeV is due to the channeling radiation 
(ChR). 
The decrease in the intensity of this peak in a periodically bent crystal (curve 2)
is associated with the decrease in the number of channeling electrons.

\begin{figure} [h]
\centering
\includegraphics[width=7.7cm,clip]{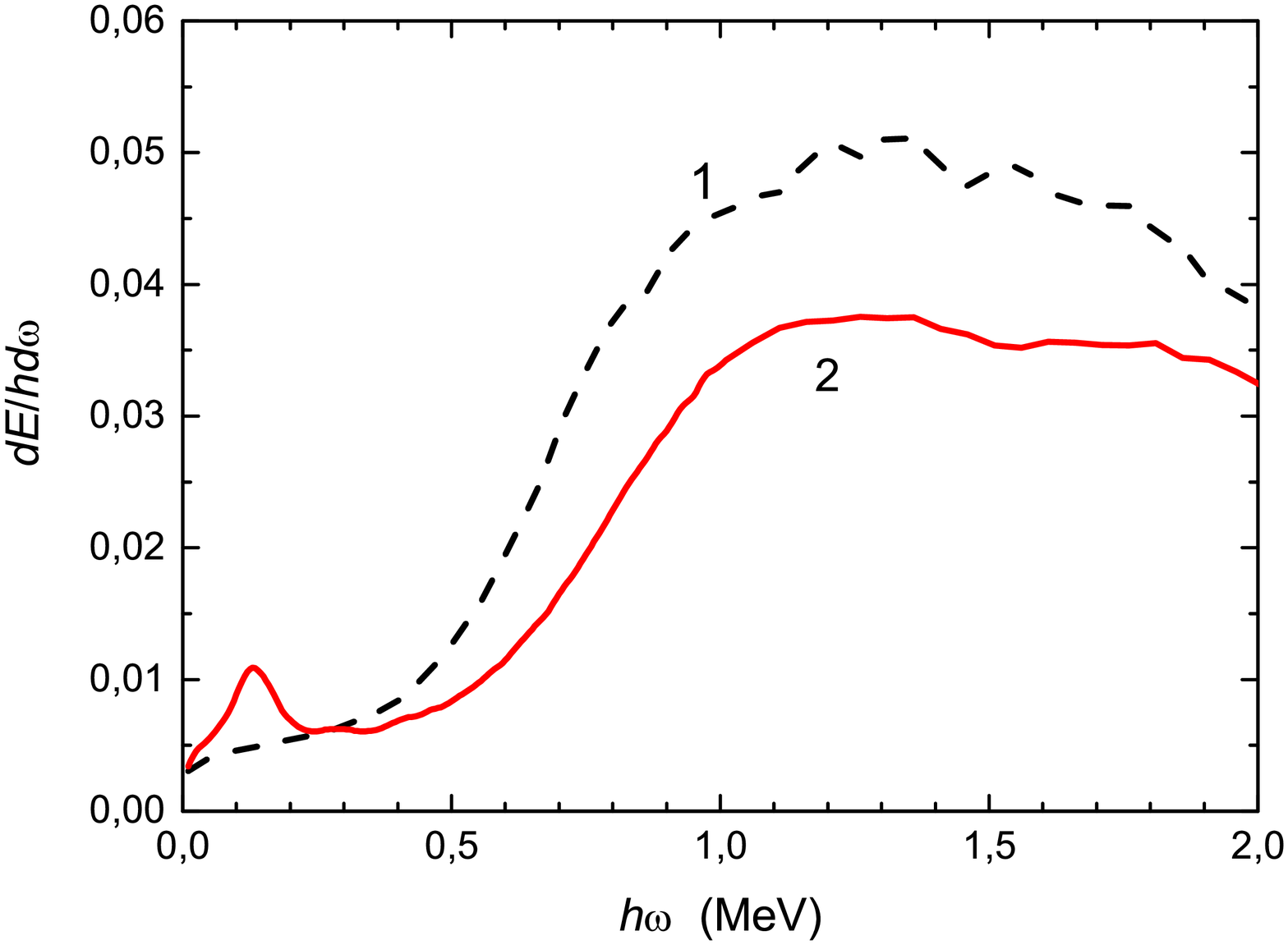}
\includegraphics[width=7.7cm,clip]{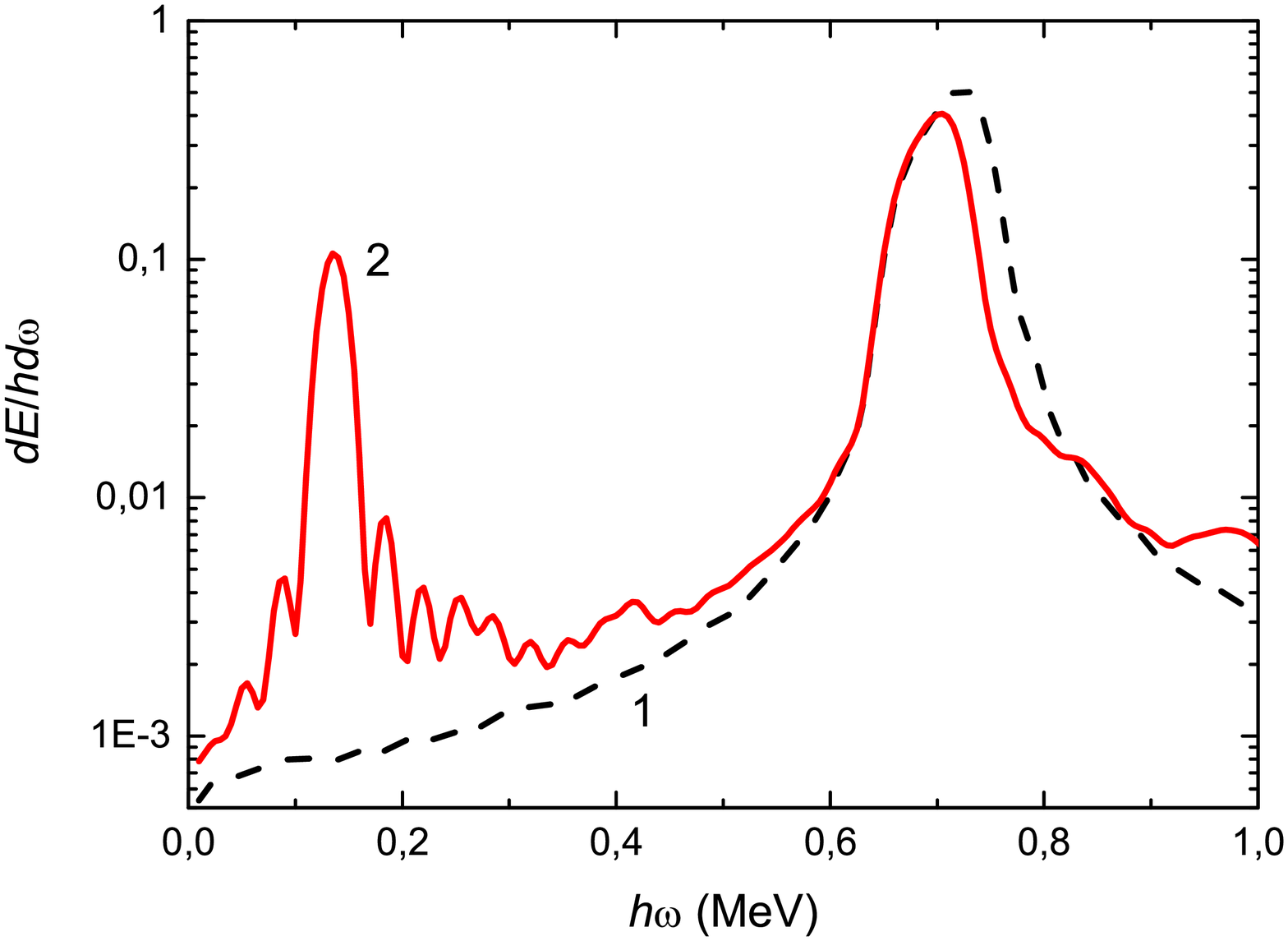}
\caption{
Representative trajectories of electrons (left) and positrons (right) with energies of 270 MeV
in a periodically bent 20 $\mu$m thick oriented diamond(110) crystal. 
Channeling (curves 1), dechanneling (2) and rechanneling (3) modes are indicated.
}
\label{CLS.fig}
\end{figure}

Figure 2 right presents the corresponding emission spectra of positrons.
Here, the ChR maximum (curve 1) is narrower and higher because the channeling oscillations
of positrons is much more harmonic that of electrons and, thus, the radiation emitted is 
concentrated in the narrower bandwidth $\Delta \om$.

It can be seen from Fig. 2 left and right (curves 2) that a radiation intensity peak is observed for
channeling in the PB crystal at a photon energy of the order of 130 keV, which is absent in
the straight crystal. 
This peak appears due to motion of channeling particles along
the centerline of the periodically bent channel. 
The particle radiation frequency is related to the period of  the channel curvature and the 
longitudinal energy of the charged particle. 
This radiation, termed as a crystalline undulator radiation, has a narrow spectral width and bears the features of 
radiation emitted by projectiles moving in magnetic undulators.
Since the study deals with electrons and positrons with the same energy, the position
of the undulator peak on the emission spectra is the same. 
However, radiation intensity is higher for positrons than for electrons by an order of magnitude, because positrons
experience harmonic oscillations and longer channeling.

\section{Conclusion}

We have numerically simulated the trajectories of ultrarelativistic charged particles
in straight and bent diamond crystals, with electrons and positrons incident on the (110)
crystallographic plane, using the MBN Explorer software package \cite{MBN_Explorer_Paper,MBN_Explorer_Site}.
The coordinates of the particles upon entering the crystal in the transverse plane were chosen with a random
number generator. 
Statistical processing of the obtained trajectories made it possible to determine the channeling 
parameters of electrons and positrons with an energy of 270 MeV in a $20$ $\mu$m thick diamond crystal.
We have established that channeled positrons have a larger acceptance and run substantially
longer distances in the crystalline channel as compared to electrons.

The calculated emission spectra of electrons and positrons channeled in a periodically bent crystal 
contain two main regions. 
The high-energy intensity peak is associated with ChR induced by oscillatory motion of the
particles in the channel; the same peak was obtained under channeling in a straight crystal.

A low-energy peak in the 130 keV region occurs when particles move in a periodically
bent channel and has an undulatory nature.
This radiation is coherent and, even though the bent crystal has a small number of periods
(only 4), the radiation is characterized by a noticeable intensity, which is significant for
potential applications in lasers \cite{KSG_review_1999,ChannelingBook2014,KSG_review2004}.

The obtained channeling parameters and the calculated emission spectra are of interest in
view of the experiments on electron channeling in straight and bent crystals currently under way
at the University of Mainz (Germany) \cite{BadEms_p58}.

\ack

This work has been supported by the European Commission (the PEARL Project within the H2020-MSCA-RISE-2015 call, GA 690991).
We acknowledge the Supercomputing Center of Saint Petersburg Polytechnic University 
(SPbPU) for providing the opportunities to carry out large-scale simulations.


\end{document}